\pdfoutput=1

\documentclass[a4paper,11pt]{article}
\usepackage{pos}

\usepackage{wrapfig}

\graphicspath{{plots/}}

\renewcommand{\logo}{\relax}

\title{Spin and 3D structure: theoretical advances}

\author*{Markus Diehl}

\affiliation{Deutsches Elektronen-Synchroton DESY,\\
  Notkestr.~85, 22607 Hamburg, Germany}

\emailAdd{markus.diehl@desy.de}

\abstract{
Theoretical studies of hadron structure have made important progress since the
previous DIS workshop in 2025.  This talk reviews selected new results and
developments, with a focus on spin dependent parton distributions, on
transverse-momentum distributions and their rapidity dependence, on generalised
parton distributions, and on the evaluation of power corrections.
}

\FullConference{33rd International Workshop on Deep Inelastic Scattering and
   Related Subjects (DIS2026)\\
   4-8 May 2026\\
   Bologna, Italy\\}

\renewcommand{\hookAfterAbstract}{%
\par\bigskip
\textsc{Preprint}: DESY--26--137 \\
}

\newcommand{\tvec}[1]{\mathbf{#1}}

\newcommand{\ccby}{\href{https://creativecommons.org/licenses/by/4.0}{CC BY 4.0}}

\newcommand{\ms}{\mskip 1.5mu}
\newcommand{\bs}{\mskip -1.5mu}

\begin{document}
\maketitle


\section{Setting the scene}

Let us recall some of the overarching questions we are addressing in the study
of hadron structure, as well as the distributions that quantify this structure
and will be discussed in this talk.

\paragraph{Hadron structure} in one longitudinal and two transverse dimensions is being studied along two distinct lines:
\begin{itemize}
\item TMDs (transverse-momentum distributions) depend on the longitudinal
momentum fraction $x$ of a parton and on its transverse momentum $\tvec{k}_T$.
This is a very natural extension of standard collinear (i.e.\ $\tvec{k}_T$
integrated) PDFs.  However --- as we shall review --- specifying $\tvec{k}_T$
has a major impact on the scale dependence of the distributions.
\item GPDs (generalised parton distributions) depend on two longitudinal
momentum fractions ($x$ and $\xi$) and on a transverse momentum transfer
$\tvec{\Delta}_T$.  The latter is Fourier conjugate to the transverse location
$\tvec{b}_T$ of the parton inside the hadron.  Sum rules, i.e.\ integrals over
$x$, connect GPDs with the form factors of local currents, in particular of the
energy-momentum tensor.
\end{itemize}
\emph{Generalised TMDs} depend on $x$, $\xi$, $\tvec{\Delta}_T$ and
$\tvec{k}_T$, combining the information of TMDs and GPDs.  Their theoretical
study is an active field, but not discussed in this talk for reasons of time.

\paragraph{Spin and orbital angular momentum} can be probed in a variety of
ways:
\begin{itemize}
\item polarised collinear PDFs form the backbone of spin studies and quantify in
particular the helicity content of a hadron at the level of its constituents.
\item polarised TMDs or GPDs can correlate the hadron or parton spin with
orbital motion, because they can depend on the cross product of vectors in
different directions.
\end{itemize}

\paragraph{Twist-three distributions} are perhaps less well known and studied,
but contain a wealth of characteristic information:
\begin{itemize}
\item they can express quantum interference, e.g.\ between scattering on a
single quark and on a quark accompanied by a gluon,
\item they are responsible for many intriguing spin phenomena, in particular for single-spin asymmetries,
\item they are closely connected with $\tvec{k}_T$ weighted integrals of certain
TMDs.
\end{itemize}

The following sections discuss \emph{selected} new developments in each of the
above areas.  More recent activity on TMDs and GPDs is mentioned on the
presentation slides.\footnote{see
\url{https://agenda.infn.it/event/47074/contributions/289180}}  No attempt has
been made to include results that appeared after the talk was given.


\section{Spin}


\subsection{Polarised PDFs}

A major step forward has been made in fits of polarised PDFs, with the first
extractions at NNLO accuracy published in 2024 and 2025:
\textsc{BDSSV24}, \textsc{MAPPDFpol1.0}, and \textsc{NNPDFpol2.0}.
This progress has been enabled by the computation of unpolarised and polarised
SIDIS cross sections at NNLO in 2023 to 2025, with independent calculations by
two groups, see \cite{Bonino:2025bqa,Goyal:2025qyu} and references therein.
This accuracy is not yet available for single-hadron or jet production in $p p$
collisions.
The recent NLO extraction \textsc{JAMpol25} \cite{Cocuzza:2025qvf} is presented
by W Melnitchouk in WG5.

\begin{figure}[bh]
\begin{center}
\includegraphics[width=0.495\textwidth]{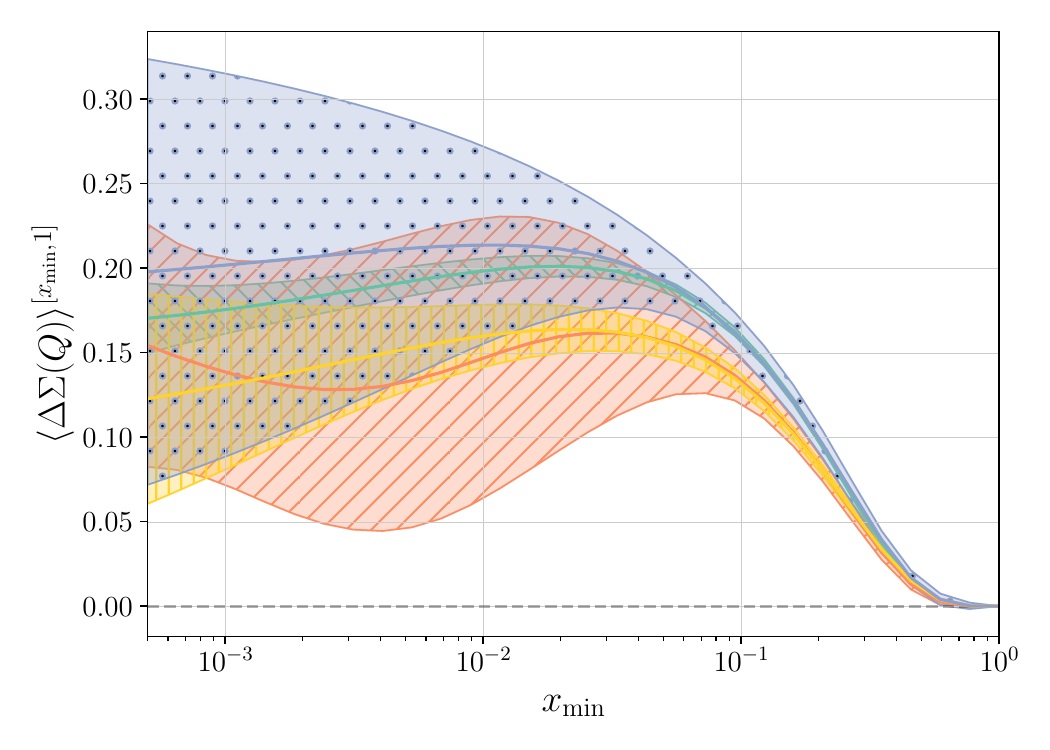}
\includegraphics[width=0.495\textwidth]{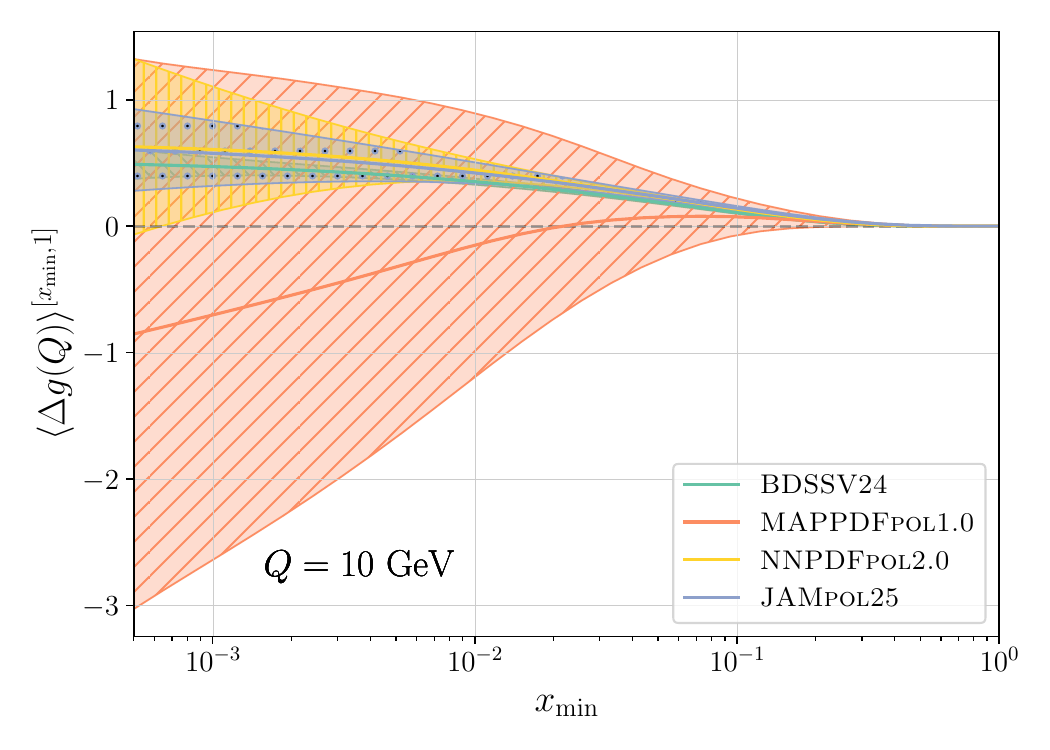}
\caption{\label{fig:spin-sr} Contributions to the proton spin of quarks and
antiquarks (left) and of gluons (right) with momentum fraction $x \ge
x_{\text{min}}$.  Plots from \protect\cite{Proceedings:2026xrb} under licence
\ccby.}
\end{center}
\end{figure}

Fig.~\ref{fig:spin-sr} from the recent recent review \cite{Proceedings:2026xrb}
shows the contributions to the proton spin of quarks and gluons with momentum
fraction $x \ge x_{\text{min}}$.  The large uncertainty on the gluon for
\textsc{MAPPDFpol1.0} reflects that no proton-proton data are included in that
analysis.  Whilst the contributions from moderate and large $x$ to the helicity
sum rule are well consolidated, the uncertainty on the helicity carried by
small-$x$ partons remains substantial.  A major step forward in this domain can
only be expected from the EIC \cite{Proceedings:2026xrb}.

Three-loop effects to inclusive DIS and to the matching of PDFs across flavour
thresholds are now fully computed, see \cite{Ablinger:2025nnq} and references
therein.  These results will be most valuable for matching the anticipated high
precision of EIC measurements, including the polarised case.


\subsection{Twist-three distributions}

Roughly speaking, parton distributions of twist three contribute to observables
that are suppressed by one power of $1 / Q$ w.r.t.\ the leading part of a cross
section; typically these are specific angular or spin asymmetries.
Twist-three distributions come in two variants, containing either two or three
parton fields in the hadronic matrix elements by which they are defined.
\begin{enumerate}
\item \emph{Two-parton distributions:} their subleading nature relative to
twist-two PDFs is due to the combination of Dirac or Lorentz indices.  Like
PDFs, these distributions depend on one momentum fraction $x$, and often they
are sufficient to describe twist-three observables at LO in $\alpha_s$.
\emph{However}, these distributions mix with three-parton distributions under
scale evolution, so they cannot be considered in isolation.  Using the equations
of motion for quark and gluon fields, one can in fact express all two-parton
distributions of twist three in terms of ordinary PDFs and of three-parton
distributions.  (This no longer holds at twist four.)
\item \emph{Three-parton distributions} depend on two independent momentum
fractions $x_1$ and $x_2$, which makes them considerably harder to model.  They
typically appear in observables starting at NLO level.  The evolution equations
governing their scale dependence are fully known at LO level.  New is that there
are now two public codes for computing this evolution numerically
\cite{Rodini:2024usc}.  They have been used for a global extraction of
twist-three distributions from a range of observables in
\cite{Vladimirov:2025qrh}; this analysis is presented by A Vladimirov in WG5.
\end{enumerate}
Fragmentation functions of twist three can be discussed along the same lines,
with the important difference that no public evolution code is available.

\begin{figure}[hb]
\begin{center}
\includegraphics[width=0.99\textwidth]{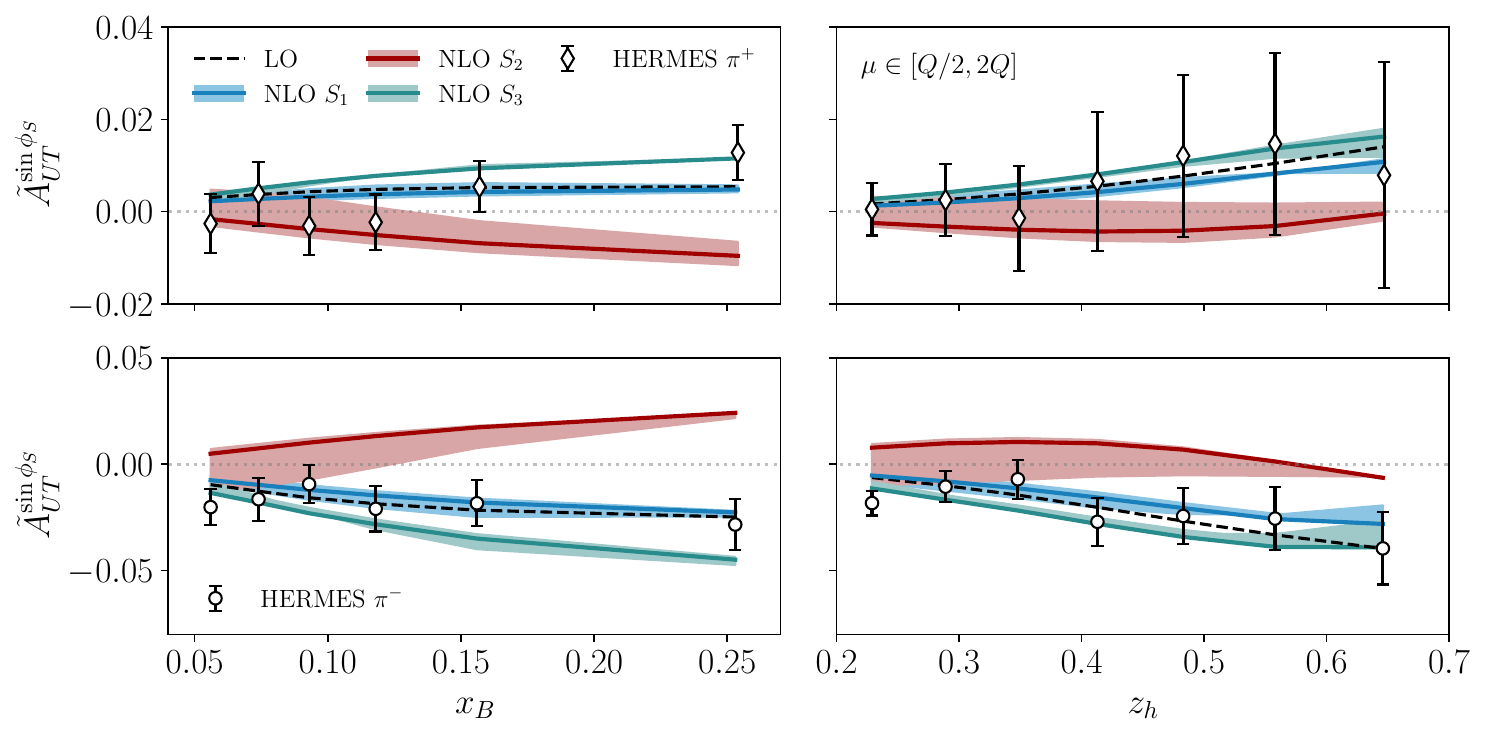}
\caption{\label{fig:tw-three} The transverse target spin asymmetry in SIDIS.
Results for three different models of the fragmentation functions entering at
NLO are denoted by $S_{1, 2, 3}$.  Plots from
\protect\cite{Scantamburlo:2026hcp} under licence \ccby.}
\end{center}
\end{figure}

A relatively simple twist-three observable is the transverse target spin
asymmetry in SIDIS, integrated over the transverse hadron momentum.  At LO, this
only involves the twist-two transversity distribution $h_1(x)$ and a twist-three
fragmentation function called $\widetilde{H}(z)$, both being chiral-odd.  The
NLO corrections that contain twist-two PDFs and twist-three fragmentation
functions were recently computed in \cite{Scantamburlo:2026hcp}.  Collinear
divergences in the hard scattering are found to cancel after adding the
counterterms corresponding to the evolution of the PDFs and fragmentation
functions; this is not trivial because there is no proof of factorisation for
this type of observable.
The study constructs different models for the three-particle fragmentation
functions entering the observable at NLO, all compatible with the extraction of
$\widetilde{H}(z)$ in \cite{Gamberg:2022kdb}. Fig.~\ref{fig:tw-three} shows that
NLO corrections can have a sizeable impact on the asymmetry and may even change
its size.  Moreover, one sees that experimental data is able to distinguish
between different model assumptions.

The recent calculation \cite{Rein:2025qhe} of a different single-spin asymmetry
in lepton-proton scattering also finds that NLO corrections may be large.  This
underlines the importance of pushing the phenomenology of twist-three
observables beyond leading order.


\section{TMDs}

The last years have seen an increasing number of extractions of unpolarised TMDs
from SIDIS and Drell-Yan data with measured transverse momentum much smaller
than the hard scale $Q$.  Recently a number of analyses have appeared that are
reduced in scope (notably by including only Drell-Yan data) but push forward the
methodology in different directions.   They may hence be regarded as proofs of
feasibility.
The analysis \cite{Bacchetta:2025ara}, presented by A Bacchetta in WG5, uses
neural networks to parametrise TMDs in the non-perturbative regime, and the work
\cite{Kang:2026mod} employs AI tools at various stages of the fitting procedure.
A joint extraction of unpolarized TMDs and PDFs was carried out in
\cite{Barry:2025glq}.  The study \cite{Avkhadiev:2025wps} is described
in the next subsection.


\subsection{TMD evolution and the Collins-Soper kernel}

In addition to the kinematical variables $x$ and $\tvec{k}_T$, TMDs depend on an
ultraviolet renormalisation scale $\mu$ and on a so-called rapidity parameter
$\zeta$, which has the dimension of mass squared.  The $\mu$ dependence has the
form
\begin{align}
   \frac{\partial f_a}{\partial \ln \mu}
   &=
   \Bigl[ \gamma_{K\bs,\ms a} \bigl(\alpha_s(\mu)\bigr) \,
            \ln\frac{\mu}{\sqrt{\zeta}}
      + \gamma_a(\alpha_s(\mu)\bigr)
   \Bigr] \,
   f_a
   \,,
   &
   a = q, g
\end{align}
with anomalous dimensions $\gamma_{K\bs,\ms a}$ and $\gamma_a$ that are spin
independent and only differ between quarks and gluons.  $\gamma_{K\bs,\ms a}$ is
known to $\mathcal{O}(\alpha_s^5)$ and $\gamma_a$ to $\mathcal{O}(\alpha_s^4)$.
Notice that the $\mu$ dependence of TMDs is very different from the one of
collinear PDFs, which is given by the DGLAP equations.

The parameter $\zeta$ specifies the rapidity range of soft gluons included in
the TMD.  The evolution in this variable is given by the \emph{Collins-Soper
equation}
\begin{align}
   \frac{\partial f_a(x, \tvec{b}_T, \mu, \zeta)}{\partial \ln \sqrt{\zeta}}
   &=
   K_a(\tvec{b}_T, \mu) \cdot f_a(x, \tvec{b}_T, \mu, \zeta)
   \,,
\end{align}
which has a simple multiplicative form after a two-dimensional Fourier transform
from $\tvec{k}_T$ to a transverse distance $\tvec{b}_T$ called ``impact
parameter''.

The physics of the Collins-Soper kernel $K_a(b_T, \mu)$ is related with soft
gluon exchange between the spectator partons of different hadrons in a physical
process like SIDIS or Drell-Yan production.  Higher $\zeta$ means more
longitudinal phase space for such soft gluon exchange.  Technically, $K_a$ is
defined via the vacuum expectation value of Wilson line operators.  This makes
it highly universal: it is independent of polarisation and hadron species and
only differs between quarks and gluons.

At small $b_T$ the Collins-Soper kernel has been computed in perturbation theory
up to $\mathcal{O}(\alpha_s^4)$.  At large $b_T$ it is a non-perturbative
quantity for which methods in lattice QCD have been developed in the last
decade.  There are several recent studies of the kernel $K_q$ for quarks
\cite{Bollweg:2025iol, Alexandrou:2025xci, Tan:2025ofx}.  Using the universality
just mentioned, they extract the kernel from TMD-type matrix elements between a
pion and the vacuum, rather than between two proton states.

\begin{figure}
\begin{center}
\includegraphics[width=0.495\textwidth]{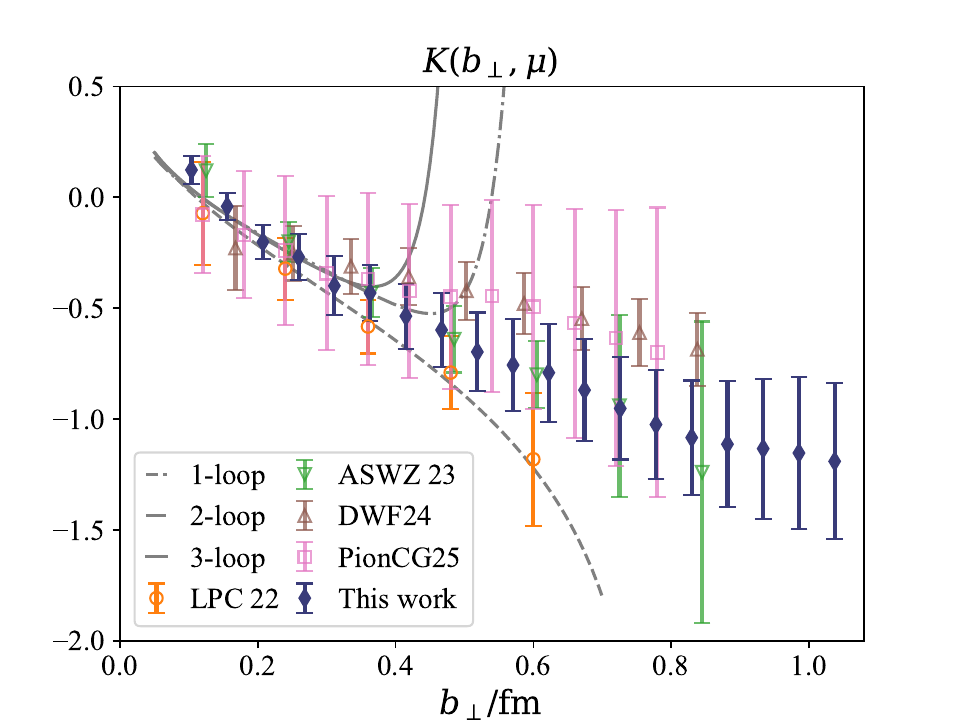}
\includegraphics[width=0.495\textwidth]{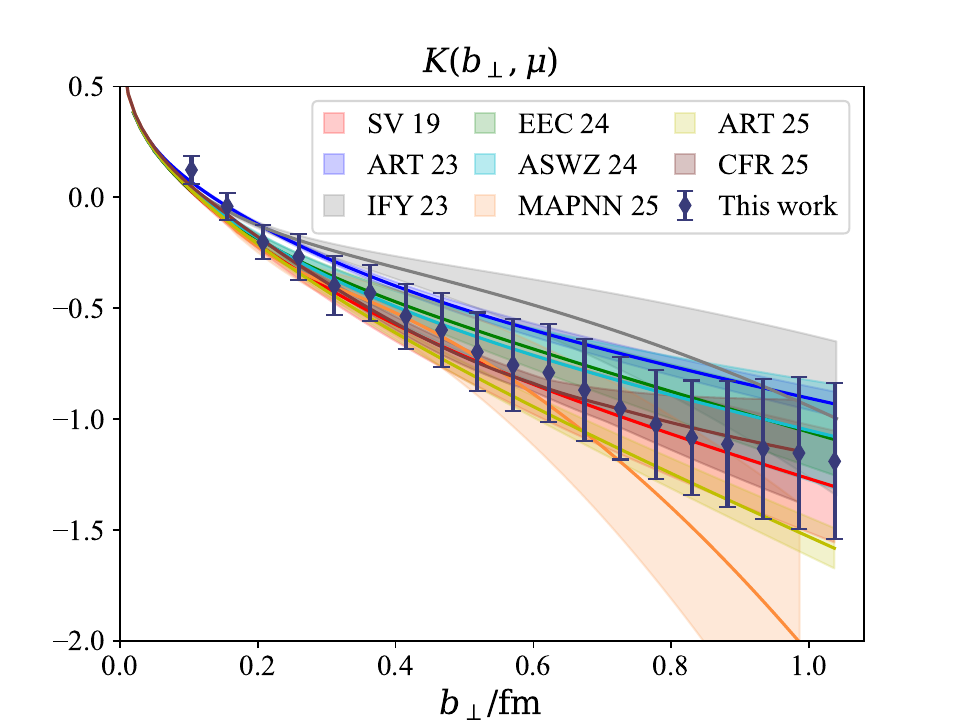}
\caption{\label{fig:cs-lat-pheno} Comparison of results for the Collins-Soper
kernel $K_q$ of quarks from lattice QCD (left) and from TMD fits to experiment
(right).  Plots from \protect\cite{Tan:2025ofx} under licence \ccby.}
\end{center}
\end{figure}

Fig.~\ref{fig:cs-lat-pheno} compares recent determinations of $K_q$ from lattice
QCD and from TMD fits.  The left panel also shows that the perturbative
calculation is valid up to distances around $0.3 \operatorname{fm}$ but not
beyond. One finds reasonable agreement between different determinations within
their uncertainties, which tend to increase with $b_T$ (denoted by $b_\perp$ in
the figure).

The negative sign of $K_q$ at large $b_T$ implies that with increasing $\zeta$,
i.e.\ with increasing phase space for soft gluon exchange, the $k_T$
distribution becomes broader.  This is quite intuitive; what is remarkable is
that with TMD factorisation this effect can be computed in terms of a single
vacuum matrix element.

The study \cite{Avkhadiev:2025wps} combines the setup of the TMD fit
\cite{Bacchetta:2025ara} with recent lattice data for the Collins-Soper kernel.
It finds that the two types of input can be combined consistently, either by
reweighting the TMD fit with the lattice results, or by fitting experimental and
lattice data together. Compared to the TMD determination without lattice input,
one observes a shift of $K_q$ by about 10\% and a decrease of its uncertainty by
about 50\%.

A similar conclusion is reached in the study \cite{Cridge:2025wwo}, presented by
T Cridge in WG1.  To assess the uncertainty on $\alpha_s$ extracted from the
$p_T$ spectrum of Drell-Yan production around the $Z$ peak, it performs a number
of fits to pseudo-data in the kinematics of the ATLAS measurement
\cite{ATLAS:2023lhg}.  The fits include up to two non-perturbative parameters
for the Collins-Soper kernel.  The resulting uncertainty on $\alpha_s$ strongly
degrades when attempting to determine these parameters from the $p_T$ spectrum
alone; including input from lattice QCD can in turn restore sensitivity
to~$\alpha_s$.


\subsection{Power corrections in SIDIS}
\label{sec:sidis-pow}

SIDIS with measured transverse hadron momentum $P_{h \perp} \ll Q$ is one of the
main observables for TMDs, along with Drell-Yan production with measured
$q_T \ll Q$ of the lepton pair.  There is significant activity in evaluating
power suppressed contributions to these processes, driven by the fact that
certain observables are zero at leading-power accuracy.  Examples are the $\cos
\phi_h$ asymmetry at order $1/Q$, and the longitudinal structure function $F_L$
or the $\cos 2\phi_h$ asymmetry at order $1/ Q^2$.  A complete factorised
description is emerging at order $1/ Q$ (Vladimirov et al.\ 2021--23; Ebert,
Gao, Stewart 2021; Gamberg et al.\ 2022), but this has not yet been used in
phenomenology. At order $1/Q^2$ it is not even known whether the cross section
factorises at all.

\begin{figure}
\begin{center}
\includegraphics[width=0.485\textwidth]{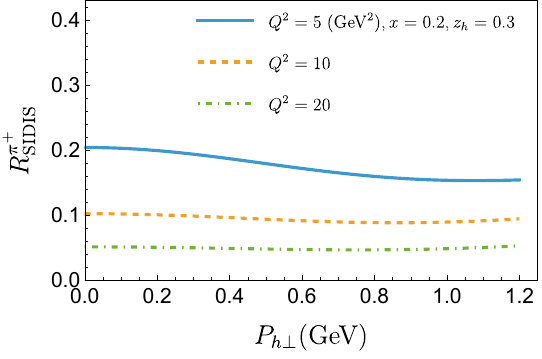}
\hspace{0.7em}
\includegraphics[width=0.485\textwidth]{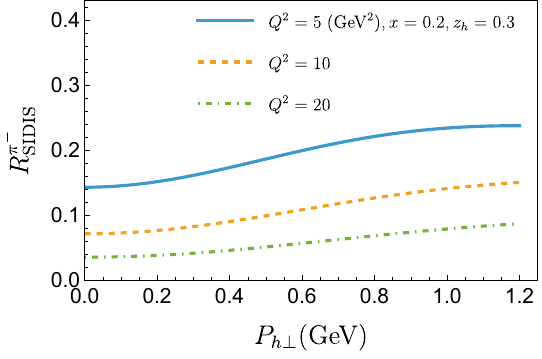}
\caption{\label{fig:R-SIDIS} Ratio of cross sections for longitudinal or
transverse photons in SIDIS producing a $\pi^+$ or a $\pi^-$. The computation
includes kinematic power corrections. Plots from \protect\cite{Balitsky:2026nux}
under licence \mbox{\ccby}.}
\end{center}
\end{figure}

New results by two groups have appeared for \emph{kinematic} power corrections
in SIDIS, which capture \emph{part} of the $1/Q$ or $1/Q^2$ suppressed
contributions and can be fully expressed in terms of twist-two TMDs
\cite{Piloneta:2025jjb, Balitsky:2026nux}.  Both approaches have previously been
applied to the Drell-Yan process.
The results of the two studies partially differ, which is expected because of
their different scope.  They concur however in the primary conclusion that
formally power suppressed terms can be sizeable in experimentally relevant
kinematics, not only for fixed-target experiments but also at the EIC.  An
example is shown in fig.~\ref{fig:R-SIDIS} for the ratio $R = F_L / F_T$ in
SIDIS.


\section{GPDs}

Whilst PDFs and TMDs appear at the cross-section level, GPDs enter directly
in the \emph{amplitudes} of suitable exclusive processes.  They depend on two
longitudinal momentum fractions $x \pm \xi$ and on the invariant momentum
transfer $t$ to the hadron.  One can trade $t$ for the transverse momentum
transfer $\tvec{\Delta}_T$.  A two-dimensional Fourier transform turns
$\tvec{\Delta}_T$ into an impact parameter $\tvec{b}_T$ that specifies the
transverse position of the selected parton in the hadron.  Note that this is
\emph{not} the same variable as $\tvec{b}_T$ defined for TMDs, which is
conjugate to a transverse parton momentum~$\tvec{k}_T$.

The scale dependence of GPDs is governed by an extended form of DGLAP
evolution, with kernels that are fully known at $\mathcal{O}(\alpha_s^2)$.
The theory description of processes involving GPDs is most advanced for Compton
scattering with at least one photon far off-shell: namely DVCS ($\gamma^*\bs p \to
\gamma p$, measured in $\ell p \to \ell \gamma p$), timelike Compton scattering
($\gamma p\to \gamma^*\bs p$, measured in $\gamma p \to \ell^+ \ell^- p$), and
double DVCS ($\gamma^*\bs p \to \gamma^*\bs p$, measured e.g.\ in $e p \to e\ms
\mu^+\bs \mu^- p$).
In all cases, the hard scattering at leading power in $1/Q$ is fully known to
NLO accuracy.  For DVCS the computation of NNLO corrections has been completed
in recent years (V Braun et al.~2020--22; Y Ji and J Sch\"onleber 2023).


\subsection{Lattice determinations}

The $x$ moments of GPDs are linear combinations of form factors of local
currents and have been computed in lattice QCD for a long time.  More recently,
the techniques of quasi- or pseudo-distributions have allowed for first lattice
evaluations of the $x$ dependence of GPDs, and recent studies \cite{Chu:2025kew,
Gao:2025inf} report results for nonzero skewness $\xi$.

\begin{figure}
\begin{center}
\includegraphics[width=0.99\textwidth,trim=0 0 0 265,clip]{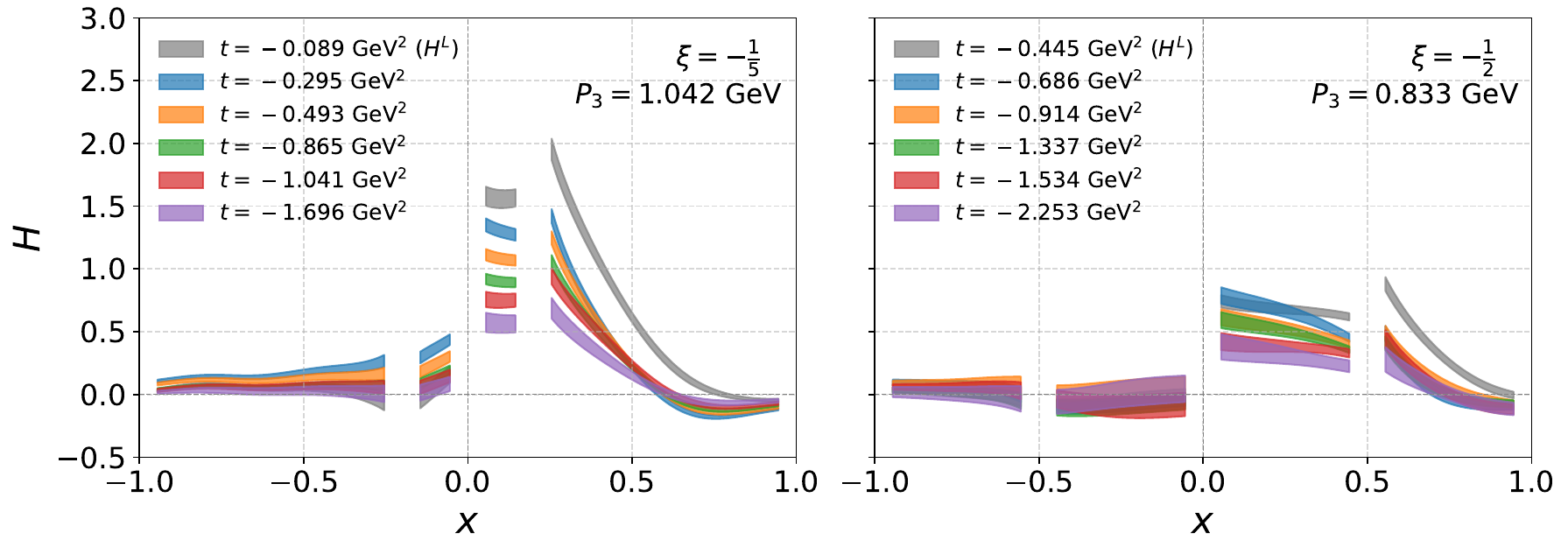}
\caption{\label{fig:gpd-lat} Lattice results for the isovector GPD $H^u - H^d$
in a proton at scale $\mu = 2 \operatorname{GeV}$.  The pion mass of the
simulation is $m_\pi \approx 260 \operatorname{MeV}$.  Plots from
\protect\cite{Chu:2025kew} under licence \ccby.}
\end{center}
\end{figure}

Fig.~\ref{fig:gpd-lat} shows results for the $x$ and $t$ dependence at two
values of $\xi$.  The white gaps around $x = 0$, $x = \pm\ms \xi$ and $x = \pm
1$ reflect that the employed method breaks down in these regions. Despite this
limitation, the plots carry an important message: the shape of the $x$
dependence changes with $t$ (and hence with $b_T$ after the appropriate Fourier
transform). The simple assumption of a factorised $t$ dependence is hence not
only disfavoured by a number of theoretical or phenomenological considerations,
but also by lattice results.


\subsection{Power corrections in DVCS}

In a similar manner as for SIDIS (sect.~\ref{sec:sidis-pow}), contributions
suppressed by powers of $1/Q$ are important for DVCS.  Certain angular
modulations only start at the power suppressed level, and even more than for
SIDIS, event rates for DVCS strongly decrease with $Q$.  Factorisation at the
first subleading order $1/Q$ has been systematically investigated in
\cite{Schoenleber:2024ihr} and includes GPDs of twist three.

Recent years have seen major progress in the evaluation of kinematic power
corrections, which can be expressed in terms of twist-two GPDs only.  In DVCS
they appear parametrically in the forms $\xi m_{\bs p} / Q$ and $\Delta_T / Q$.
The latter is especially important, because the resolution in $b_T$ space
scales like the inverse of the largest $\Delta_T$ that can be measured
\emph{and} analysed in the GPD framework.

\begin{figure}[hb]
\begin{center}
\includegraphics[width=0.495\textwidth]{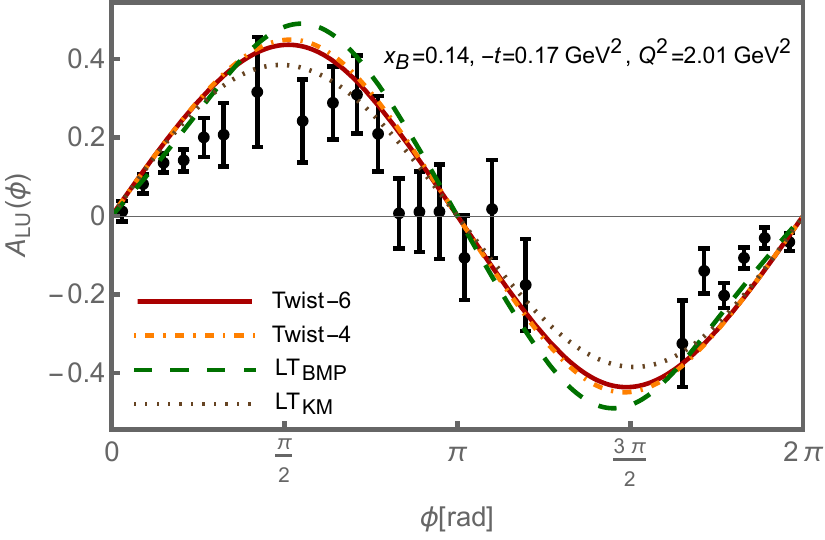}
\includegraphics[width=0.495\textwidth]{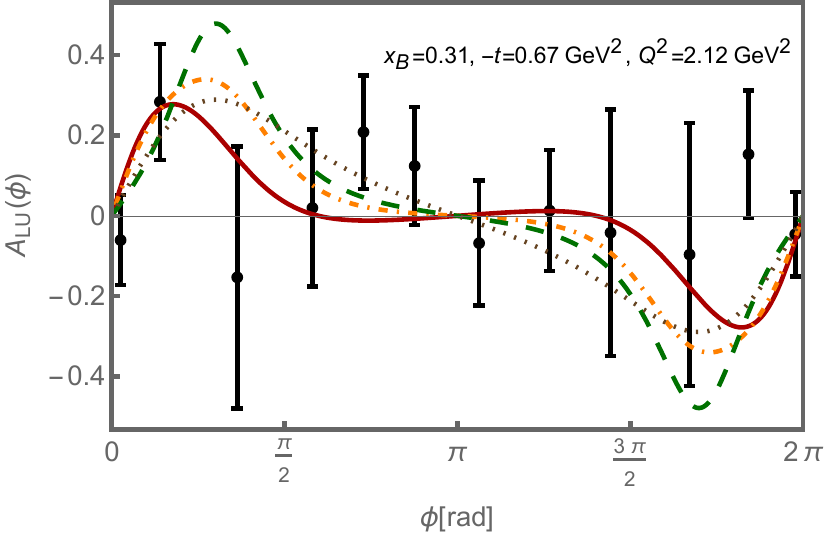}
\caption{\label{fig:dvcs-pow} The angular dependence of the beam spin asymmetry
in DVCS, computed with kinematic power corrections.  The association between
twist and powers is: Twist~6 $\leftrightarrow 1/Q^4$, Twist~4 $\leftrightarrow
1/Q^2$, and \mbox{LT $\leftrightarrow 1/Q^0$}.  Plots from
\protect\cite{Braun:2025xlp} under licence \ccby.}
\end{center}
\end{figure}

The recent work \cite{Braun:2025xlp} presents kinematic power corrections up to
order $1/Q^4$ for all helicity amplitudes in a spin-$1/2$ target.  This allows
one to study the stability of the power expansion in~$1/Q$.  In numerical
estimates, the authors find that this expansion is reasonably stable for $|t|
\lesssim Q^2 / 4$, but not much beyond. This is illustrated in
fig.~\ref{fig:dvcs-pow}.

\section{Summary}

Considerable theory progress has been made in all areas reviewed in this talk.
Fits of longitudinally polarised collinear PDFs have entered the high-precision
area, building on an increasing number of NNLO results.  For the determination
of twist-three distributions, NLO corrections can be essential to obtain stable
theory predictions.

Fits of unpolarized TMDs are becoming increasingly sophisticated, driven to a
good part by the highly precise LHC data for the $p_T$ spectrum in $Z$
production.  By contrast, the phenomenology of polarised TMDs typically remains
limited by the statistical precision of available data.  The EIC is expected to
change this radically.

For both TMDs and GPDs new lattice methods are being developed, and their
results can have a significant impact on phenomenology.  New theoretical
approaches allow for the evaluation of kinematic power corrections in SIDIS and
in DVCS, and estimates indicate that these corrections can be sizeable in
experimentally relevant kinematics.  This raises the important question how
large power corrections of non-kinematic origin may be in such situations.  At
twist-three level, theoretical frameworks are being developed to address this
question, but more work is needed to turn these into practice.


\acknowledgments

It is a pleasure to thank J Bl\"umlein, V Braun, Y Ji, G Marinelli, M Schlegel,
F Tackmann, A~Vladimirov, W~Vogelsang, and F Wunder for helpful input or
discussions.


\bibliography{spin_3d_thy.bib}

\end{document}